\documentclass[
    final            
  ]
  {aipproc}

\layoutstyle{6x9}

\begin{document}

\title{Band structure of boron doped carbon nanotubes}

\author{Ludger Wirtz}{
address={Department of Material Physics, University of the Basque
Country, Centro Mixto CSIC-UPV, and Donostia International Physics Center,
Po.~Manuel de Lardizabal 4, \\
20018 Donostia-San Sebasti\'an, Spain}
}

\author{Angel Rubio}{
address={Department of Material Physics, University of the Basque
Country, Centro Mixto CSIC-UPV, and Donostia International Physics Center,
Po.~Manuel de Lardizabal 4, \\
20018 Donostia-San Sebasti\'an, Spain}
}

\begin{abstract}
We present {\it ab initio} and self-consistent tight-binding calculations
on the band structure of single wall semiconducting carbon nanotubes
with high degrees (up to 25 \%) of boron substitution. Besides a lowering of
the Fermi energy into the valence band, a regular, periodic distribution
of the p-dopants leads to the formation of a dispersive ``acceptor''-like band 
in the band gap of the undoped tube.
This comes from the superposition of acceptor levels at the boron 
atoms with the delocalized carbon $\pi$-orbitals.
Irregular (random) boron-doping leads to a high concentration of hybrids
of acceptor and unoccupied carbon states above the Fermi edge.
\end{abstract}

\maketitle


\section{Introduction}
The electronic properties of single wall carbon nanotubes 
depend sensitively on the diameter and the chirality of the
tubes. Therefore, in order to use tubes as elements in nano-electronical
devices, a controlled way to produce and separate a large
quantity of tubes of specific radius and chirality has to be
found. Alternatively, doping of tubes by boron and nitrogen \cite{doping}
may lead to electronic properties that are more controlled by the
chemistry (i.e., the amount of doping) than the specific geometry of the tubes.
Indeed, theoretical investigations of BCN nanotube heterojunctions have
predicted that the characteristics of these junctions are largely
independent of geometrical parameters \cite{blas97}.
It has been predicted that even stochastic doping may lead to 
useful electronic elements such as chains of random quantum dots
and nanoscale diodes in series \cite{lam01}. 

The realization of p-type doping of pure carbon nanotubes by a substitution reaction
with boron atoms \cite{han99,gol99,bor03} offered the possibility
to transform semiconducting tubes into metallic tubes by lowering
the Fermi level into the valence band.
Indeed, transport measurements \cite{ter98,wei99,liu01} have shown
a clearly enhanced conductivity of B-doped tubes. The metallic behavior
of B-doped multiwall carbon nanotubes was also confirmed by scanning tunneling
spectroscopy \cite{car98}.

In bulk semiconductors, typical doping concentrations of impurity 
atoms are around $10^{-3}$ \%. At such low concentration, the presence
of group III atoms (e.g., B) in a group IV semiconductor 
leads to the formation of an acceptor state (non-dispersive band) in the
band gap at low energy above the valence band edge \cite{ashcroft}. 
For B doping in a C(8,0) tube, Yi and Bernholc \cite{yi93}
have calculated (using a B/C ratio of 1/80) that this acceptor
state is located at 0.16 eV above the Fermi energy. However no discussion has been made
up-to-date on the evolution of this acceptor-like level with the degree of B-doping. 
Since in recent experiments \cite{bor03}, Boron substitution
up to 15 \% has been reported, we investigate in this paper
the band structure of strongly B-doped single wall carbon nanotubes.

\section{Computational Method}
In order to demonstrate the effect of B doping on a semiconducting tube, 
we have chosen a (16,0) tube with a diameter of 12.5 {\AA} which is close
to the average radius of commonly produced SWNT samples. 
Since the distribution of borons in B-doped tubes is still unknown,
we have performed two sets of calculations. 
1.) For regular periodic distributions that are commensurate with the unit-cell 
of the undoped tubes, we perform {\it ab initio} calculations 
\cite{footabinit} of electronic band structure and density of states (DOS). 
2.) In order to test the effect of disordered B doping,  
we employ a supercell comprising 6 unit-cells (384 atoms) and distribute the
boron atoms ``randomly'' with the only constraint that they cannot
occupy nearest neighbor positions on the hexagonal carbon grid.
For this large system, we use a self-consistent tight-binding (SCTB) 
method \cite{lam01,footsctb}.

\begin{figure}
  \includegraphics[width=1.\textwidth]{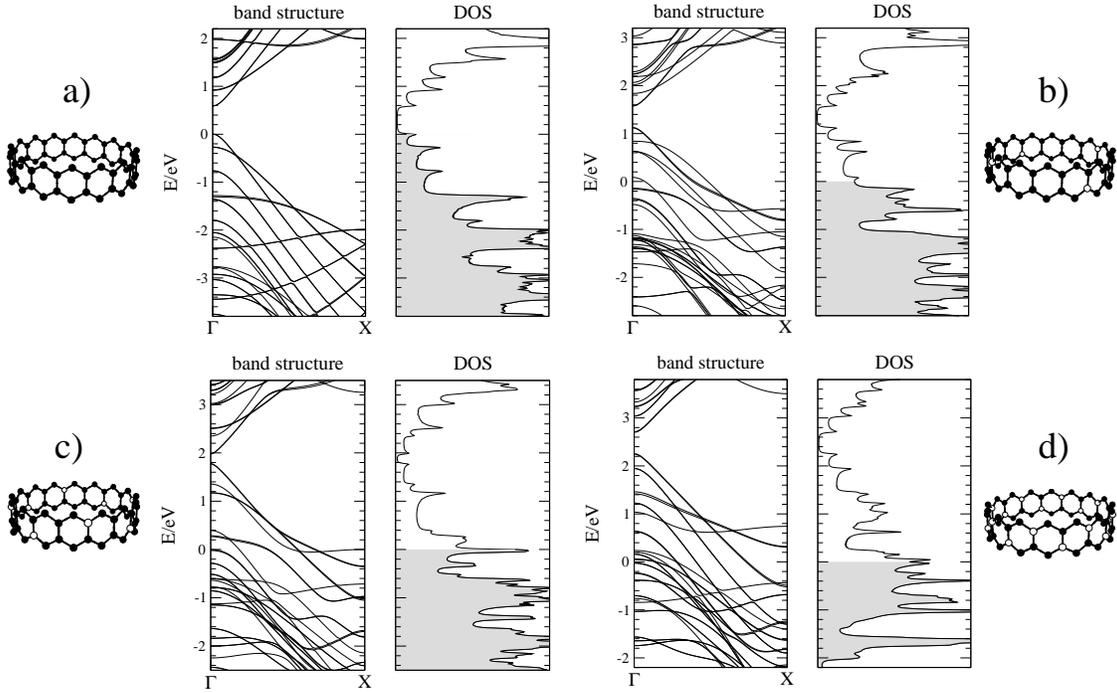}
  \caption{Band structure and DOS of a C(16,0) tube with a) 0\%, b) 6.25\%,
  c) 12.5\%, and d) 25\% boron doping. Zero energy denotes the Fermi edge.
  Filled states are indicated by grey shadowing. Insets display the
  corresponding unit cells (C atoms black, B atoms white).}
\end{figure}

\section{Results}
\begin{figure}
  \includegraphics[width=0.65\textwidth]{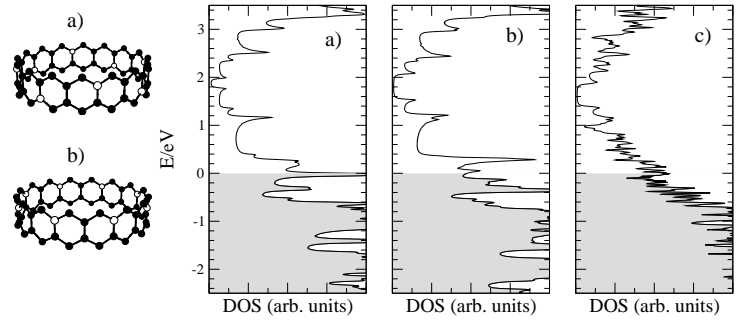}
  \caption{DOS of a C(16,0) tube with 12.5\% B doping: 
  a) regular doping (as in Fig.~1), b) regular doping (alternative
  geometry, c) random doping.}
\end{figure}
Fig. 1 shows the band structure and DOS of a
C(16,0) tube without doping and with various concentrations of B-doping
in a periodic manner (see insets). For the undoped tube it can be
noted, that the DOS is symmetric around the band gap for the 1st and 2nd 
Van-Hove singularities (VHSs). However, beyond this small energy regime around
the band gap, asymmetries arise due to the mixing of p and s orbitals. 
Doping of the tubes leads to a lowering of the Fermi level into the 
valence band of the undoped tube. Clearly, the stronger the doping,
the stronger is also the shift of the Fermi level.
Above the highest occupied bands of the undoped tube new bands are formed.
These bands correspond to the formation of an acceptor level in semiconductors
with very low dopant concentration. Due to the strong concentration
in the present case, the boron levels hybridize with the carbon levels
and form strongly dispersive ``acceptor-like bands''. The band structure of
the valence band is strongly distorted upon doping, because the
levels of carbon atoms that are substituted by boron move upwards in
energy. In addition, the lowering of the symmetry by doping leads to many clear
avoided crossings between states that perform a real crossing in the
case of the undoped tube where they posses a different symmetry.
Clearly, in all cases displayed in Fig.~1, the doping with B leads to
a metallic character of the tube. However, the density of states at the
Fermi level strongly depends on the geometry of the structure and varies
non-monotonously
with the doping concentration. In fact, in BC$_3$ tubes with a different 
geometry than in Fig.~1 d), a gap between $\pi$-type orbitals and $\sigma$ 
type orbitals opens up at the Fermi level and renders the tubes 
semiconducting in this particular case \cite{miy94}.

Fig.~2 compares the DOS of a C(16,0) tube with 12.5\% boron substitution
in three different geometries. The two regular structures both display
a pattern with the familiar pronounced Van Hove singularities of a 1-dim.
band structure. However, the different ordering of the boron atoms gives rise 
to a different hybridization of the bands and thereby to a pronounced
difference in the peak structure of the DOS. The DOS at the
Fermi level strongly depends on the ordering of the B atoms.
In the DOS of the randomly doped tube, the pattern of the VHSs is 
mostly smeared out \cite{foot}. Due to the missing 
translational geometry, the 1-dim. band structure of
the regularly doped tubes is transformed into the DOS of a 0-dimensional
structure, i.e. of a large molecule or cluster without periodicity.
The states between the Fermi level and the original valence band edge
of the undoped tubes are hybrids of acceptor levels and unoccupied
carbon levels.
The electronic excitations into these levels should explain
the optical absorption spectra of B-doped carbon tubes \cite{bor03}: 
In addition to the pronounced absorption peaks that are commonly affiliated
with the transitions $E_{ii}$ ($i=1,2$) from the first/second 
occupied VHS to the 
first/second unoccupied VHS of the pure semiconducting carbon tubes, 
the spectra of B-doped tubes display additional absorption at energies 
lower than $E_{11}$.

\section{Conclusion}
The calculations on the band structure of boron doped carbon nanotubes
clearly confirm the expectation (and experimental observation) that these
tubes are metallic with low resistance. For regularly doped structures,
we have observed the formation of a dispersive ``acceptor'' band while
the Fermi level is shifted downwards into the valence band of the
undoped tubes. Electronic excitations into these hybrids of acceptor states
and unoccupied carbon levels are expected to play an important role in optical 
absorption spectra.
Randomly doped tubes display the same downshift of
the Fermi edge but cease to display strongly dispersive bands.
We hope that in the near future, spatially resolved TEM/EELS will help
to elucidate the exact geometry of the boron dopants and 
scanning tunneling spectroscopy will probe the exact density
of states.


Work supported by COMELCAN (contract number HPRN-CT-2000-00128).
We acknowledge stimulating discussion with T. Pichler, J. Fink
and G. G. Fuentes.

\bibliographystyle{aipproc}   

\begin{thebibliography}{1}
\bibitem{foot} The system still possesses a translational symmetry even though
with a large unit cell. Therefore, some of the dispersive band structure, 
including Van-Hove singularities, is still visible.
\end{thebibliography}

\end{document}